\documentclass[twocolumn,showpacs,preprintnumbers,amsmath,amssymb]{revtex4}

\usepackage{graphicx}
\usepackage{dcolumn}
\usepackage{bm}
\usepackage{bbm}

\begin{document}

\preprint{APS/123-QED}

\title{Atomic quantum superposition state generation via optical probing}

\author{Anne E. B. Nielsen}
\author{Uffe V. Poulsen}
\author{Antonio Negretti}
\altaffiliation{Present address: Institute for Quantum Information Processing, University of Ulm, Albert-Einstein-Allee 11, D-89069 Ulm, Germany.}
\author{Klaus M{\o}lmer}
\affiliation{Lundbeck Foundation Theoretical Center for Quantum
System Research, Department of Physics and Astronomy,
Aarhus University, DK-8000 \AA rhus C, Denmark}

\date{\today}

\begin{abstract}
We analyze the performance of a protocol to prepare an atomic ensemble in a superposition of two macroscopically distinguishable states. The protocol relies on conditional measurements performed on a light field, which interacts with the atoms inside an optical cavity prior to detection, and we investigate cavity enhanced probing with continuous beams of both coherent and squeezed light. The stochastic master equations used in the analysis are expressed in terms of the Hamiltonian of the probed system and the interaction between the probed system and the probe field and are thus quite generally applicable.
\end{abstract}

\pacs{42.50.Dv, 42.50.Pq}
\keywords{Suggested keywords}

\maketitle

\section{Introduction}

The ability to prepare quantum mechanical systems in superposition states is important both in fundamental studies of quantum mechanics and in technological applications. Generation of such superposition states of light by conditioning on the outcome of a measurement has been achieved experimentally \cite{neergaard,grangier,sasaki}, but light fields are difficult to store for long periods of time, and it is desirable to be able to prepare trapped atoms in superposition states as well. First steps in this direction have demonstrated generation of superposition states of up to 6 trapped ions \cite{leibfried} and $W$ states of up to 8 trapped ions \cite{haffner} through interactions with suitably chosen light pulses. In the present paper we consider trapped atomic ensembles and investigate the possibility to prepare superposition states by means of optical quantum non-demolition measurements.

The idea to probe the state of an atomic system by allowing the system to interact with a light field and then subject the light field to measurements is very useful. It has, for instance, been used to observe quantum jumps of single ions \cite{qjump1,qjump2,qjump3}. Another application is the generation of spin squeezed states by means of quantum non-demolition measurements \cite{kuzmich}, and related proposals use spin squeezing to improve the precision of atomic clocks \cite{meiser,appel} and magnetometers \cite{budker}. Several recent experiments \cite{chapman,fortier,hinds,esslinger,reichel,gupta,brennecke} have focused on the possibility to trap cold atoms and Bose-Einstein condensates inside high-finesse optical cavities, because the cavity enhances the light-atom interaction strength, and the strong-coupling regime, where the coherent dynamics takes place on a faster time scale than the dissipative dynamics, has been reached. In addition, optical measurements can be used to investigate the atom statistics of atom laser beams \cite{ottl} and the atomic population statistics of optical lattices \cite{ritsch2,ritsch,eckert}.

In the following, we demonstrate that for the strong-coupling parameters obtained in \cite{reichel}, a measurement procedure similar to the quantum non-demolition measurement used to generate spin squeezed states can be used to generate quantum superposition states, as outlined in \cite{nm6} (see also \cite{polzik} for a related proposal). In Sec.~\ref{II} we explain the state preparation protocol, and we provide the time evolution of the state of the atoms and the light field. The performance of the protocol is analyzed in Sec.~\ref{III}.

It is convenient to use a continuum coherent state from a laser as the probe field, but it is also interesting to investigate the additional possibilities that arise, if we choose to probe the system with different quantum mechanical states of light. An example of free-space probing of an atomic ensemble with single-mode photon subtracted squeezed vacuum states has been investigated in \cite{fiurasek}, and in Sec.~\ref{IV} of the present paper we provide general tools to analyze probing with a continuous beam of squeezed light. Despite the infinite dimensionality of the Hilbert space of the light field it is possible, for the case at hand, to rewrite the equation governing the time evolution of the state of the atoms and the light field into a finite set of coupled stochastic differential equations, and this allows us to obtain numerical results for the full dynamics. Section \ref{V} concludes the paper.

\section{Probing procedure}\label{II}

\begin{figure}
\includegraphics*[width=0.92\columnwidth]{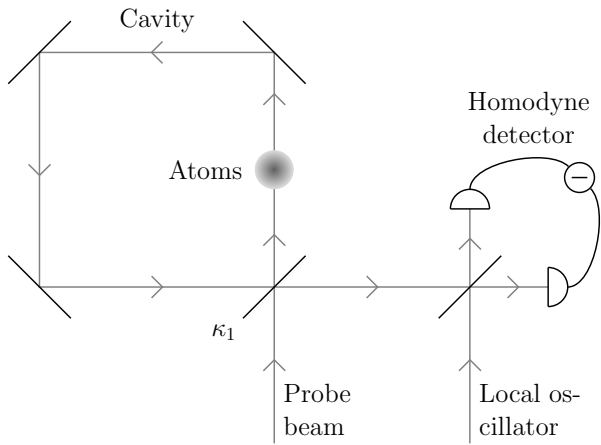}
\caption{The atoms are placed inside an optical cavity and are probed with a light beam, which is in a continuum coherent state before it enters into the cavity. The probe light reflected from the cavity is observed with a homodyne detector and the atomic quantum superposition state is generated for certain possible measurement readouts.\label{setup}}
\end{figure}

To probe atoms inside an optical cavity, we send a probe beam, which is initially in a continuum coherent state, into the cavity and observe the light reflected from the cavity with a homodyne detector as shown in Fig.~\ref{setup}. We consider atoms with two ground state levels, which we treat as a spin-1/2 system, and we assume that the cavity field couples one of these levels non-resonantly to an excited state. Assuming that all atoms couple identically to the cavity field, which may be achieved e.g.\ with a Bose-Einstein condensate, and eliminating the excited state adiabatically, the Hamiltonian for the light-atom interaction takes the form
\begin{equation}\label{H}
H=\hbar\tilde{g}\hat{a}^\dag\hat{a}\hat{J}_z,
\end{equation}
where $\tilde{g}\equiv g^2/\Delta$, $g$ is the single-atom coupling strength on the optical transition, $\Delta=\omega-\omega_0$ is the detuning, $\omega$ is the frequency of the light field, $\omega_0$ is the frequency of the atomic transition, $\hat{a}$ is the field annihilation operator of the cavity field, $\hat{J}_z$ is the $z$-component of the collective atomic spin vector ${\bm \hat{J}}\equiv\sum_i{\bm \hat{j}}_i$, and ${\bm \hat{j}}_i$ is the spin vector of the $i$th atom. We have here ignored the possibility of spontaneous emission from the excited state, which is valid in the strong coupling limit. We return to this point in Sec.~\ref{III}.

To understand why superpositions of very different quantum states are obtained for certain measurement outcomes, if the phase of the local oscillator is chosen appropriately, we first consider the case, where the initial state of the atoms is a single eigenstate of $\hat{J}_z$ with eigenvalue $\hbar n$. In this case the unitary time evolution operator corresponding to the above Hamiltonian reduces to a phase shift operator, and provided the cavity field is in the vacuum state at time $t=0$, one finds that the cavity field at time $t$ is a coherent state with amplitude
\begin{equation}\label{alphan}
\alpha_n(t)=\sqrt{\kappa_1}\int_0^te^{-(\kappa/2+in\tilde{g})(t-t')}\beta(t')dt',
\end{equation}
where $\kappa_1$ is the cavity decay rate due to the input mirror,
$\kappa=\kappa_1+\kappa_{\textrm{loss},1}$ is the total cavity decay
rate including loss, and $\beta(t')$ is the amplitude of the probe
beam, i.e., $|\beta(t)|^2\Delta t$ is the expectation value of the
number of photons in the beam segment of length $c\Delta t$, which
arrives at the input mirror at time $t$. Note that \eqref{alphan} is
independent of the actual outcome of the homodyne detection, which
appears because the quadrature noise of the coherent state leaving the
cavity is uncorrelated with the quadrature noise of the coherent state
inside the cavity. Choosing $\beta(t')$ to be real, it
follows from \eqref{alphan} that $\alpha_n(t)=\alpha_{-n}^*(t)$, and a
measurement of the $x$-quadrature of the cavity field is thus unable
to distinguish between $n$ and $-n$. This suggest that for a general
initial state of the atoms, the homodyne detection slowly projects the
atomic state onto a superposition of states with the same value of
$|n|$.

\begin{figure}
\includegraphics*[width=\columnwidth]{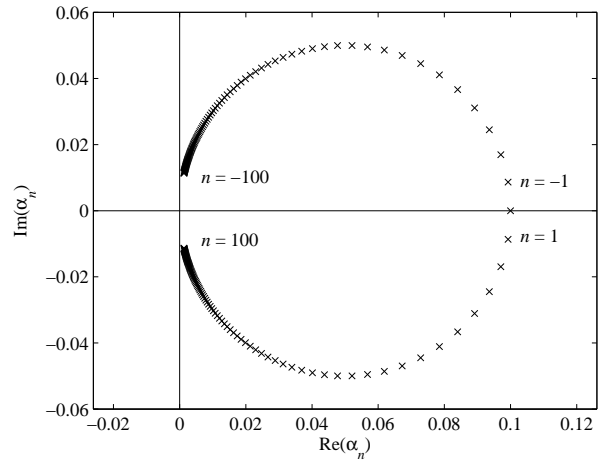}
\caption{Real and imaginary parts of $\alpha_n$ for $g=2\pi\times215$~MHz, $\Delta=2\pi\times10$~GHz, $\kappa_1=\kappa=2\pi\times106$~MHz, $4\kappa_1\beta^2/\kappa^2=0.01$, and $n=-100$, $-99$, $\ldots$, $100$. The points lie on a circle with center $(\sqrt{\kappa_1}\beta/\kappa,0)$ and radius $\sqrt{\kappa_1}\beta/\kappa$.\label{alphaplot}}
\end{figure}

If we choose $\beta(t')$ to be time independent (in a frame rotating
with the relevant bare cavity resonance frequency), $\alpha_n(t)$
assumes the steady state value
\begin{equation}\label{alphanss}
\alpha_n=\frac{2\sqrt{\kappa_1}\beta}{\kappa}
\frac{1-2in\tilde{g}/\kappa}{1+4n^2\tilde{g}^2/\kappa^2}
\end{equation}
after a transient time of order a few $\kappa^{-1}$. The amplitude $\alpha_n$ is plotted in Fig.~\ref{alphaplot} for the values of $g$ and $\kappa$ in Ref.~\cite{reichel} and different values of $n$. For small
$2|n|\tilde{g}/\kappa$, $\alpha_n$ is simply rotated by an angle
proportional to $n$ compared to $\alpha_0$, but when the rotation
angle increases, the amplitude of the cavity field drops,
because the probe field is no longer on resonance due to the phase
shift introduced by the atoms, and as $2|n|\tilde{g}/\kappa$ tends to
infinity, $\alpha_n$ approaches the origin. Thus, $|\alpha_n-\alpha_{-n}|$
is small for $2|n|\tilde{g}/\kappa\gg1$, which means that the
rate of decoherence between $|n\rangle$ and $|-n\rangle$ due to cavity
losses and detection inefficiency is reduced. To illustrate this
explicitly we note that if the initial state of the atoms is an equal
and pure superposition of $n$ and $-n$, then it follows from
Eq.~\eqref{rho} below that the purity of the atomic state evolves as
\begin{multline}
\textrm{Tr}\left(\rho_{\textrm{at}}^2(t)\right)=\frac{1}{2}\bigg(1+
\exp\bigg\{-\frac{4\kappa_1\beta^2n^2\tilde{g}^2}
{[(\kappa/2)^2+n^2\tilde{g}^2]^2}\\
\times\left[(\kappa-\eta\kappa_1)t+1\right]\bigg\}\bigg),
\end{multline}
where $\rho_{\textrm{at}}(t)$ is the atomic density operator obtained
by tracing out the cavity field, $\eta$ is the detection efficiency,
and we have neglected transient terms proportional to $\exp(-\kappa t/2)$ and to $\exp(-\kappa t)$ in the exponent. In addition, the smaller number of
photons in the cavity for large $2|n|\tilde{g}/\kappa$ leads to a
decrease in the rate of spontaneous emission from the excited atomic
state.

The stochastic master equation for the setup in Fig.~\ref{setup} was
derived in Ref.\ \cite{nm6} for $n\tilde{g}\tau\ll1$,
$\kappa\tau\ll1$, and $\sqrt{\kappa_1}|\beta|\tau\ll1$, where $\tau$
is the round trip time of light in the cavity, and these requirements
are all fulfilled for the parameters in Fig.~\ref{alphaplot} and a
cavity length of $38.6\textrm{ }\mu\textrm{m}$, as in
Ref.~\cite{reichel}. In the following, we consider the case, where the
initial state of the atoms is a coherent spin state pointing in the
$x$-direction, and for notational simplicity, we thus restrict
ourselves to a basis consisting of simultaneous eigenstates of ${\bm \hat{J}}^2$ and $\hat{J}_z$ with total spin quantum number $J=N/2$, where $N$ is the number of atoms. This is possible because the Hamiltonian commutes with ${\bm\hat{J}}^2$. If we also assume $\beta=\beta^*$ and a local oscillator phase, which corresponds to a measurement of the $x$-quadrature of the cavity field, the solution takes the form
\begin{widetext}
\begin{multline}\label{rho}
  \rho(t)=\sum_{n=-J}^J\sum_{m=-J}^JC_{nm}(0)\exp\bigg[-\frac{\kappa}{2}\int_0^t\big(|\alpha_n(t')|^2
  +|\alpha_m(t')|^2-2\alpha_n(t')\alpha_m^*(t')\big)dt'
  +\sqrt{\eta\kappa_1}\int_0^t\left(\alpha_n(t')+\alpha_m^*(t')\right)dy'\\
  -\frac{1}{2}\eta\kappa_1\int_0^t\left(\alpha_n(t')+\alpha_m^*(t')\right)^2dt'
  -\frac{1}{2}\sqrt{\kappa_1}\int_0^t\beta(t')\left(\alpha_n(t')-\alpha_n^*(t')
    -\alpha_m(t')+\alpha_m^*(t')\right)dt'\bigg]\\
  |\alpha_n(t)\rangle\langle\alpha_m(t)|\otimes|n\rangle\langle m|
  \bigg/\sum_{q=-J}^JC_{qq}(0)\exp\left[\sqrt{\eta\kappa_1}\int_0^t(\alpha_q(t')+\alpha_q^*(t'))dy'-
    \frac{1}{2}\eta\kappa_1\int_0^t(\alpha_q(t')+\alpha_q^*(t'))^2dt'\right],
\end{multline}
\end{widetext}
where $\rho(t)$ is the density operator describing the state of the
cavity field and the atoms, $|\alpha_n(t)\rangle$ is a coherent state
with amplitude \eqref{alphan}, $|n\rangle$ is the eigenstate of $\hat{J}_z$
with eigenvalue $\hbar n$, and $dy'$ is proportional to the
photocurrent measured in the time interval from $t'$ to $t'+dt'$ (see
Ref.\ \cite{nm6} for details). The coefficients $C_{nm}(0)$,
normalized according to $\sum_qC_{qq}(0)=1$, are determined from the
initial state of the atoms, and for a coherent spin state pointing in
the $x$-direction
\begin{equation}\label{Cnm0}
C_{nm}(0)=\frac{1}{4^J}\frac{(2J)!}{\sqrt{(J+n)!(J-n)!(J+m)!(J-m)!}}.
\end{equation}
We note that the trace over the atomic state of $\rho(t)$ is an
incoherent sum of the states $|\alpha_n\rangle\langle\alpha_n|$, and
if the different coherent states are given appropriate weights,
Fig.~\ref{alphaplot} may be regarded as a phase space representation
of the cavity field. If, instead, we trace out the light field to
obtain $\rho_{\textrm{at}}(t)$, we observe that the weights of
$|n\rangle\langle n|$ and $|-n\rangle\langle -n|$ are equal at all
times if they are equal at the initial time, as expected. Note also
that the time evolution of the diagonal elements of
$\rho_{\textrm{at}}(t)$ is independent of the off-diagonal elements of
$\rho_{\textrm{at}}(t)$.

The density operator at time $t$ depends on the results of the
continuous measurement, and the probability density to observe a given
sequence $dy_1$, $dy_2$, $\ldots$, $dy_M$ of measurement outcomes in
the interval from $0$ to $t$ is
\begin{multline}\label{PM}
P(dy_1,dy_2,\ldots,dy_M)=\prod_{k=1}^M\left[\frac{1}{\sqrt{2\pi dt}}\exp\left(-\frac{dy_k^2}{2dt}\right)\right]\\
\times\sum_{q=-J}^JC_{qq}(0)
\exp\bigg\{2\sqrt{\eta\kappa_1}\sum_{k=1}^M\textrm{Re}[\alpha_q((k-1)dt)]dy_k\\
-2\eta\kappa_1\int_0^t[\textrm{Re}(\alpha_q(t'))]^2dt'\bigg\},
\end{multline}
where $dt\equiv t/M$ and $M$ is very large. One could now use Eq.~\eqref{PM} to derive the joint probability density for the $2(N+1)$ stochastic variables $\int_0^t\textrm{Re}(\alpha_n(t'))dy'$ and $\int_0^t\textrm{Im}(\alpha_n(t'))dy'$, but for constant $\beta$ and a total probing time, which is large compared to $\kappa^{-1}$, we may approximate $\alpha_n(t')$ by its steady state value $\alpha_n$. In this case, $\rho(t)$ only depends on the stochastic variable $Y\equiv \int_0^tdy'$, which has probability density
\begin{equation}\label{PY}
P(Y)=\sum_{q=-J}^J\frac{C_{qq}(0)}{\sqrt{2\pi t}}
\exp\left[-\frac{(Y-2\sqrt{\eta\kappa_1}\textrm{Re}(\alpha_q)t)^2}{2t}\right].
\end{equation}

\section{Performance of the state preparation protocol}\label{III}

\begin{figure}
{\flushleft{(a)}\\}
\includegraphics*[width=\columnwidth]{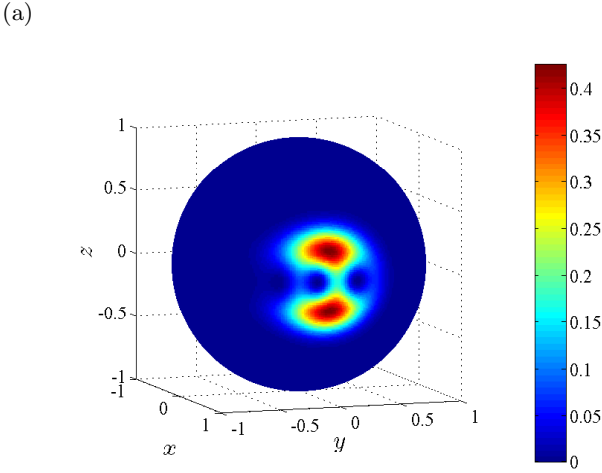}\\
{\flushleft{(b)}\\}
\includegraphics*[width=\columnwidth]{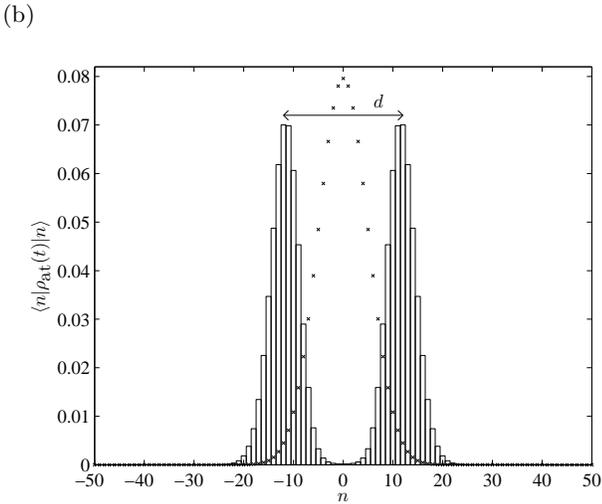}
\caption{(Color online) Spin Q-function \cite{arecchi,spinQ} on the Bloch sphere (a) and diagonal of the atomic density operator (b) for the state produced at $t=10^{-6}$~s, assuming $Y=5\times10^{-4}\textrm{ s}^{1/2}$, $J=50$, $\eta=1$, and otherwise the same parameters as in Fig.~\ref{alphaplot}. The crosses in (b) indicate the diagonal of the atomic density operator at the initial time.\label{state}}
\end{figure}

The ability to distinguish the atomic states $|n\rangle$
and $|m\rangle$ is determined by the distance
$|\textrm{Re}(\alpha_n-\alpha_m)|$ between the corresponding coherent
states measured along the $x$-axis in phase space. A glance at
Fig.~\ref{alphaplot} thus reveals that we may conditionally produce
states, for which the distribution $\langle
n|\rho_{\textrm{at}}(t)|n\rangle$ consists of two peaks separated by a
region with $\langle n|\rho_{\textrm{at}}(t)|n\rangle\approx0$, after
a relatively short probing time, while it takes significantly longer
to produce a superposition of states with a single value of $|n|$. To
estimate the required probing time to distinguish between $|0\rangle$
and $|\pm J\rangle$, we note that the probability density \eqref{PY}
is a sum of $2J+1$ Gaussians with mean values
$2\sqrt{\eta\kappa_1}\textrm{Re}(\alpha_q)t$ and standard deviations
$\sqrt{t}$. For $4\tilde{g}^2J^2/\kappa^2\gg1$,
$\textrm{Re}(\alpha_{\pm J})\approx0$ (see Eq.~\eqref{alphanss}), and the Gaussians corresponding to $n=0$ and $n=\pm J$ thus begin to separate after a probing time
\begin{equation}\label{t}
  t_{\textrm{qs}}\approx\frac{1}{\eta\kappa_1}\frac{\kappa^2}{4\kappa_1\beta^2},
\end{equation}
which evaluates to 150 ns for the parameters in Fig.~\ref{alphaplot} and $\eta=1$. To clearly distinguish between $n=0$ and $n=\pm J$, we need to choose a probing time, which is somewhat larger than this, and Fig.~\ref{state} suggests that $t\approx10^{-6}$~s is a reasonable choice for the considered parameters. Note that $t_{\textrm{qs}}$ is independent of the number of atoms as long as $4\tilde{g}^2J^2/\kappa^2\gg1$. Note also that $t\approx10^{-6}$~s is large compared to $\kappa^{-1}=1.5\times10^{-9}$~s, which justifies the steady state assumption $\kappa t\gg1$. The maximal probing time is limited by decay processes, and, in particular, we require that $t$ is small compared to the average time $t_{\textrm{sp}}$ between spontaneous emission events, which may be approximated as the inverse of the product of the atomic decay rate $\Gamma$, the probability for a single atom to be in the excited state, and the number of atoms in the state, which interacts with the light field, i.e.,
\begin{equation}\label{tsp}
t_{\textrm{sp}}\approx\left[\Gamma\frac{4\kappa_1\beta^2}{\kappa^2}
\left(1+\frac{4g^4\tilde{n}^2}{\Delta^2\kappa^2}\right)^{-1}
\frac{g^2}{\Delta^2+(\Gamma/2)^2}\frac{N}{2}\right]^{-1},
\end{equation}
where $\tilde{n}$ is a typical value of $n$. The smallest approximate value of $t_{\textrm{sp}}$ is obtained for $\tilde{n}=0$ and is $6\times10^{-5}$~s for the parameters in Fig.~\ref{alphaplot} and $\Gamma=2\pi\times6$~MHz (as in Ref.~\cite{reichel}), while we obtain $t_{\textrm{sp}}\approx2\times10^{-4}$~s if we choose $|\tilde{n}|$ to be the most probable value of $|n|$ after a probing time of $t=10^{-6}$~s and a measurement outcome $Y=0\textrm{ s}^{1/2}$ (see Eq.~\eqref{np} below). It is thus realistic to finish the probing before one of the atoms decay by spontaneous emission, and we hence neglect atomic decay throughout. Note that this conclusion is independent of $\beta$ since the rate of gain of information and the rate of spontaneous emission events are both proportional to the probe beam intensity.

\begin{figure}
\includegraphics*[width=\columnwidth]{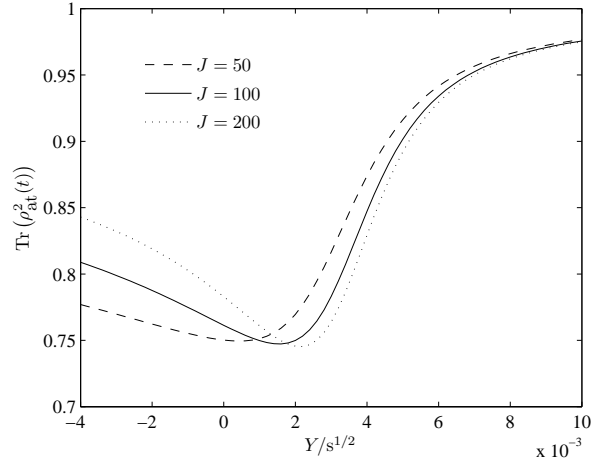}
\caption{Purity of the atomic state as a function of the integrated photo current $Y$ for $t=10^{-6}$~s, $\eta\kappa_1=0.9~\kappa$, $g=2\pi\times215$~MHz, $\Delta=2\pi\times10$~GHz, $\kappa=2\pi\times106$~MHz, and $4\kappa_1\beta^2/\kappa^2=0.01$. \label{purity3}}
\end{figure}

\begin{figure}
\includegraphics*[width=\columnwidth]{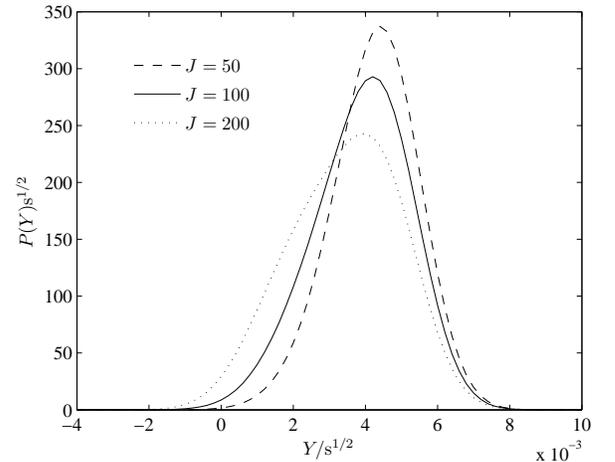}
\caption{Probability density to measure a given value of $Y$ for the same parameters as in Fig.~\ref{purity3}. When the number of atoms increases the probability to observe lower values of $Y$ increases because extra coherent states with small amplitude are added to the left in Fig.~\ref{alphaplot} and the width of the distribution $\langle n|\rho_{\textrm{at}}(0)|n\rangle$ at the initial time is broader.\label{probability}}
\end{figure}

An important measure of the quality of the generated atomic states is the purity
\begin{widetext}
\begin{multline}
\textrm{Tr}\left(\rho_{\textrm{at}}^2(t)\right)=\sum_{n=-J}^J\sum_{m=-J}^J|C_{nm}(0)|^2
\exp\bigg[-|\alpha_n(t)-\alpha_m(t)|^2-(\kappa-\eta\kappa_1)\int_0^t|\alpha_n(t')-\alpha_m(t')|^2dt'\\
+\sqrt{\eta\kappa_1}\int_0^t(\alpha_n(t')+\alpha_n^*(t')+\alpha_m(t')+\alpha_m^*(t'))dy'
-\frac{\eta\kappa_1}{2}\int_0^t((\alpha_n(t')+\alpha_n^*(t'))^2+(\alpha_m(t')+\alpha_m^*(t'))^2)dt'\bigg]\\
\bigg/\bigg\{\sum_{q=-J}^JC_{qq}(0)\exp\bigg[\sqrt{\eta\kappa_1}\int_0^t(\alpha_q(t')+\alpha_q^*(t'))dy'
-\frac{\eta\kappa_1}{2}\int_0^t(\alpha_q(t')+\alpha_q^*(t'))^2dt'\bigg]\bigg\}^2.
\end{multline}
\end{widetext}
The factor $\exp(-|\alpha_n(t)-\alpha_m(t)|^2)$ appears, because the cavity field is traced out to obtain the atomic state, but this reduction in purity may be eliminated by turning off the probe beam after the desired probing time, whereby the cavity field quickly decays to the vacuum state and the density operator factorizes. If, in addition, the initial state is pure, and there are no losses ($\eta\kappa_1=\kappa$), the atoms are seen to end up in a pure state, as expected. The purity is plotted as a function of the integrated photo current $Y$ for $\eta\kappa_1=0.9\textrm{ }\kappa$ in Fig.~\ref{purity3}, neglecting the transient of $\alpha_n(t)$, and the relevant range of $Y$-values may be read off from Fig.~\ref{probability}. For large positive values of $Y$, the purity is close to unity, because the atomic state $|0\rangle\langle 0|$ has a large weight. The purity is smaller for intermediate values of $Y$, because states with $\textrm{Re}(\alpha_n)\approx\textrm{Re}(\alpha_0)/2$ have large weights, and the distance in phase space between $\alpha_n$ and $\alpha_{-n}$ is large for these states. For low values of $Y$ the purity increases, because the states with large weights are closer to the origin, and $|\alpha_n-\alpha_{-n}|$ is thus smaller. The purity for low $Y$ is different for different numbers of atoms, because $|\alpha_J-\alpha_{-J}|$ depends on $J$. If the probe field is turned off at some point to ensure that $\alpha_n(t)=0$ at the final time, it follows from \eqref{rho} that the atomic state only depends on $\beta$ through $\beta^2dt$ and $\beta dy$ (note that $\alpha_n$ is proportional to $\beta$), and a scaling of $\beta$ thus corresponds to a scaling of the probing time and the measured photo current.

\begin{figure}
\includegraphics*[width=\columnwidth]{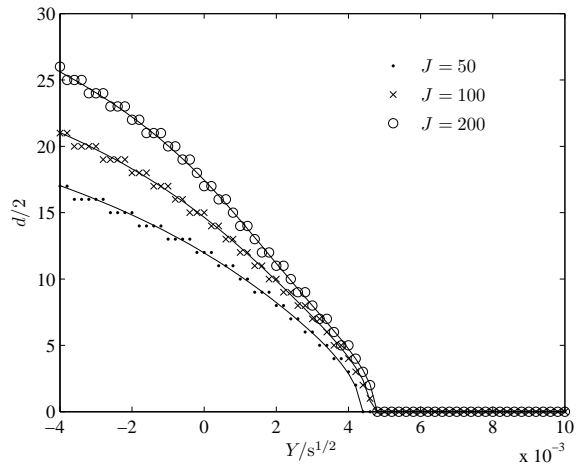}
\caption{Value of $|n|$, which maximizes $\langle n|\rho_{\textrm{at}}(t)|n\rangle$, as a function of $Y$ for the same parameters as in Fig.~\ref{purity3}. The points are obtained directly from \eqref{rho}, while the lines represent the approximation \eqref{np1}.\label{nmax}}
\end{figure}

The distance $d$ between the two peaks in Fig.~\ref{state}(b) may be used as a measure of the difference between the two states in the generated quantum superposition and is thus a second parameter to characterize the quality of the states. For large $N$ we may approximate \eqref{Cnm0} by
\begin{equation}
C_{nm}(0)=\sqrt{\frac{2}{\pi N}}\exp\left(-\frac{n^2+m^2}{N}\right),
\end{equation}
and it thus follows from Eq.~\eqref{rho} that the largest value of $\langle n|\rho_{\textrm{at}}(t)|n\rangle$ is obtained for $n=n_p$, where
\begin{equation}\label{np1}
1+\frac{bn_p^2}{N}=\frac{\sqrt{a}\left[\left(3\sqrt{3a}bt+b\sqrt{bY^3+27at^2}\right)^{2/3}-bY\right]}
{\sqrt{3}\left(3\sqrt{3a}bt+b\sqrt{bY^3+27at^2}\right)^{1/3}},
\end{equation}
and we have defined
\begin{equation}
a\equiv\sqrt{\eta\kappa_1}\times\frac{2\sqrt{\kappa_1}\beta}{\kappa}
\quad\textrm{and}\quad b\equiv\frac{4\tilde{g}^2N}{\kappa^2}.
\end{equation}
We plot $n_p$ in Fig.~\ref{nmax}, and it is seen that $n_p$ increases with decreasing $Y$, because small values of $Y$ favor atomic states with $|n|$ close to $J$. For $Y=0$ we obtain the particularly simple result
\begin{equation}\label{np}
1+\frac{4\tilde{g}^2n_p^2}{\kappa^2}=\left(2\eta\kappa_1\frac{4\kappa_1\beta^2}{\kappa^2}
\frac{4\tilde{g}^2N}{\kappa^2}t\right)^{1/3},
\end{equation}
and, in this case, $d$ scales as $N^{1/6}$ for large $N$, which means that the relative distance $d/N$ between the peaks decreases with $N$. Note, however, that the probability to measure $Y=0$ is different for different values of $J$. The peak-to-peak distance $d$ may be increased, by decreasing $4\tilde{g}^2/\kappa^2$, but since this moves the coherent states in Fig.~\ref{alphaplot} to the right around the edge of the circle, this will also decrease the final purity if losses are present.

\section{Probing with squeezed light}\label{IV}

Considering the conclusions from the previous sections, we would like to increase the similarity between the cavity field obtained for atomic states with $n\approx J$ and the cavity field obtained for $n\approx-J$, in order
to decrease the decoherence rate due to light field losses, and, at the same time, increase the difference between the cavity field obtained for atomic states with $n\approx0$ and the cavity field obtained for $n\approx\pm J$, in order to decrease the required probing time. The former may be accomplished by setting $\beta=0$, but this choice also prevents us from retrieving any information about the state of the atoms from the measurement records. If, on the other hand, we squeeze the probe beam before it enters into the cavity containing the atoms, the cavity field is different for different values of $n$ even if $\beta=0$. This suggests that it might be possible to increase the purity of the final atomic states by using a squeezed vacuum probe field, and we thus investigate this possibility in the following. For the sake of generality, we shall, however, not put $\beta=0$ until Sec.~\ref{sec:num_results_squeeze}.

\subsection{Equation of motion}
\label{sec:eom}
\begin{figure}
\includegraphics*[width=0.92\columnwidth]{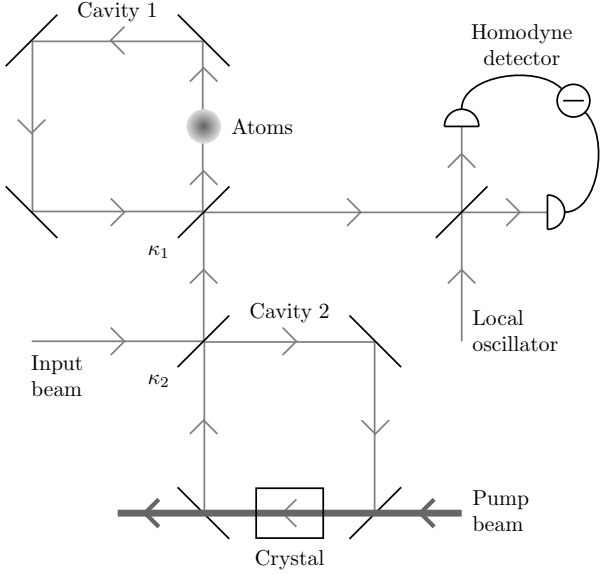}
\caption{Probing with a squeezed state may be achieved by including an optical parametric oscillator in the setup. The input beam may be either a continuum coherent state or vacuum, and the light field is squeezed when it passes through the pumped nonlinear crystal.\label{sqsetup}}
\end{figure}

A setup to probe the atoms with squeezed light is depicted in
Fig.~\ref{sqsetup} and basically consists in passing the probe beam
through a cavity containing a pumped nonlinear medium. Combining the theory developed in Refs.~\cite{nm6} and \cite{nm4}, we derive the stochastic master equation
\begin{widetext}
\begin{multline}\label{sqrho}
  \rho(t+dt)=\rho(t)-\frac{i}{\hbar}[H,\rho(t)]dt-\frac{i}{2}
  \left[(\epsilon(\hat{c}^\dag)^2+\epsilon^*\hat{c}^2),\rho(t)\right]dt
  +\frac{1}{2}(\kappa_1+\kappa_{\textrm{loss},1})
  \left(-\hat{a}^\dag\hat{a}\rho(t)-\rho(t)\hat{a}^\dag\hat{a}
    +2\hat{a}\rho(t)\hat{a}^\dag\right)dt\\
  +\frac{1}{2}(\kappa_2+\kappa_{\textrm{loss},2})
  \left(-\hat{c}^\dag\hat{c}\rho(t)-\rho(t)\hat{c}^\dag\hat{c}
    +2\hat{c}\rho(t)\hat{c}^\dag\right)dt
  +i\sqrt{\kappa_1\kappa_2}(\hat{a}^\dag\hat{c}\rho(t)-\rho(t)\hat{c}^\dag\hat{a}
  +\hat{a}\rho(t)\hat{c}^\dag-\hat{c}\rho(t)\hat{a}^\dag)dt\\
  +i\sqrt{\kappa_1}\beta[\hat{a}^\dag,\rho(t)]dt+i\sqrt{\kappa_1}\beta^*[\hat{a},\rho(t)]dt
  -\sqrt{\kappa_2}\beta[\hat{c}^\dag,\rho(t)]dt+\sqrt{\kappa_2}\beta^*[\hat{c},\rho(t)]dt\\
  +\Big\{-e^{-i\phi}\sqrt{\eta\kappa_1}\left[\hat{a}\rho(t)
    -\textrm{Tr}\left(\hat{a}\rho(t)\right)\rho(t)\right]
  -e^{i\phi}\sqrt{\eta\kappa_1}\left[\rho(t)\hat{a}^\dag
    -\textrm{Tr}\left(\rho(t)\hat{a}^\dag\right)\rho(t)\right]\\
  +ie^{-i\phi}\sqrt{\eta\kappa_2}\left[\hat{c}\rho(t)
    -\textrm{Tr}\left(\hat{c}\rho(t)\right)\rho(t)\right]
  -ie^{i\phi}\sqrt{\eta\kappa_2}\left[\rho(t)\hat{c}^\dag
    -\textrm{Tr}\left(\rho(t)\hat{c}^\dag\right)\rho(t)\right]\Big\}dW.
\end{multline}
\end{widetext}
Here, $\rho(t)$ is the density operator representing the state of the atoms, the light field in cavity 1 (annihilation operator $\hat{a}$), and the light field in cavity 2 (annihilation operator $\hat{c}$). (We assume that each
mirror gives rise to a phase shift of $\pi/2$ on the reflected field, and the phase of the light field thus assumes four different values along the length of each cavity. The operators $\hat{a}$ and $\hat{c}$ refer to the field passing
through the atomic cloud and the crystal, respectively.) $H$ is the
Hamiltonian for the light-atom interaction and the internal dynamics
of the atoms, $\epsilon$ is the nonlinear gain coefficient in the
crystal (the time evolution operator corresponding to one round trip
in the cavity is
$U_{\textrm{sq}}=\exp\{-i[\epsilon^*\hat{c}^2+\epsilon(\hat{c}^\dag)^2]\tau_2/2\}$,
where $\tau_2$ is the round trip time of light in cavity 2, and
$|\epsilon|/(\kappa_2+\kappa_{\textrm{loss},2})=1/2$ at threshold),
$\kappa_1$ and $\kappa_2$ are cavity decay rates due to the cavity
input mirrors as indicated in the figure, $\kappa_{\textrm{loss},1}$
and $\kappa_{\textrm{loss},2}$ are additional cavity decay rates due
to cavity losses, $\beta$ is the amplitude of the input beam, $\phi$
is the phase of the local oscillator, and $dW$ is a stochastic
variable with a Gaussian probability distribution with mean value zero
and variance $dt$.

\subsection{Symmetries}
\label{sec:time-reversal}
In order to produce symmetric superposition states of the atoms, we should be able to choose the parameters such that our detection gives information on $|n|$, but \emph{not} on the sign of $n$. Although intuitively clear from the
physical setup, to formally prove that this is possible, we consider a variation of the \emph{time-reversal operator} \cite{schiff}: Let $\hat{T}$ be an anti-unitary operator which acts on the atom and light field operators as follows~\footnote{
The difference to the conventional time-reversal is simply the
minus sign in the transformation equation for $\hat{c}$ which, in turn, is
only a consequence of the chosen phase convention.
}:
\begin{gather}
  \label{eq:def_Theta}
  \hat{T} {\bm\hat{J}} \hat{T}^{-1}=-{\bm\hat{J}}, \quad
  \hat{T} \hat{a} \hat{T}^{-1} = \hat{a}, \quad
  \hat{T} \hat{c} \hat{T}^{-1} = -\hat{c}.
\end{gather}
Writing $dW=\xi dt$ and $\rho(t+dt)=\rho(t)+\mathcal{L}_{\xi}[\rho(t)]dt$, where the subscript $\xi$ emphasizes the explicit dependence on the random measurement outcome, we then have
\begin{equation}
  \label{eq:sym}
  \hat{T}\rho(t+dt)\hat{T}^{-1}
  =\hat{T}\rho(t)\hat{T}^{-1}+
  \mathcal{L}_{\xi}\left[\hat{T}\rho(t)\hat{T}^{-1}\right]dt
\end{equation}
\emph{provided} that $\phi=k\pi$, $k\in\mathbbm{Z}$ (corresponding to
detection of the $x_1$ quadrature) and that $\beta$ and $\epsilon$ are
both purely imaginary (note the phase shift imposed on the input beam due to the presence of the squeezing cavity). Because of (\ref{eq:sym}), time-reversal symmetry is preserved during the propagation for each realization of
the noise. In addition, $H$ of Eq.~(\ref{H}) is obviously invariant
under arbitrary rotations around the $\hat{J}_z$ spin axis. Now, the spin
coherent states along $x$ that we use as initial states are not
time-reversal eigenstates, but are eigenstates of the product of
time-reversal and a rotation by $\pi$ around $\hat{J}_z$, $\exp(-i\pi
\hat{J}_z)\hat{T}$ corresponding to a reflection in the $xy$-plane of the
spin. Since $\hat{T}$ and $\exp(-i\pi\hat{J}_z)\hat{T}$ are both symmetries of the propagation, our initial state evolves into a state with the same reflection symmetry between $n$ and $-n$.

\subsection{Solution of the equation of motion}
\label{sec:gauss}

For the Hamiltonian \eqref{H}, which does not by itself give rise to transitions between the different $|n\rangle$ states, we may write the solution of \eqref{sqrho} as
\begin{equation}\label{split}
  \rho(t)=\sum_{n=-J}^J\sum_{m=-J}^J\rho_{nm}(t)|n\rangle\langle m|,
\end{equation}
where $\rho_{nm}(t)$ are operators acting on light fields in both
cavities, i.e., on a two-mode system. In the following, we use a mixed Wigner function density operator representation and define $W_{nm}(x_1,p_1,x_2,p_2,t)$ as the (two-mode) Wigner transform of $\rho_{nm}(t)$:
\begin{multline}
W_{nm}(x_1,p_1,x_2,p_2,t)=\frac{1}{4\pi^4}\iiiint\\
\textrm{Tr}\left(e^{(\eta_r+i\eta_i)\hat{a}^\dag-(\eta_r-i\eta_i)\hat{a}+
(\zeta_r+i\zeta_i)\hat{c}^\dag-(\zeta_r-i\zeta_i)\hat{c}}\rho_{nm}(t)\right)\\
e^{-\sqrt{2}i\eta_ix_1+\sqrt{2}i\eta_rp_1-\sqrt{2}i\zeta_ix_2+\sqrt{2}i\zeta_rp_2}
d\eta_rd\eta_id\zeta_rd\zeta_i.
\end{multline}
At the same time, we also need to translate the light field operators in Eq.~\eqref{sqrho} into operators acting on Wigner transforms (see \cite{gardiner} chapter 4). For a coherent input beam
\begin{multline}\label{solution}
W_{nm}(y,t)=\frac{N_{nm}(t)}{\pi^2\sqrt{\det(V_{nm}(t))}}\times\\
  \exp\big[
  -(y-y_{nm}(t))^T(V_{nm}(t))^{-1}(y-y_{nm}(t))
  \big],
\end{multline}
where $y=(x_1,p_1,x_2,p_2)^T$ is a column vector of the quadrature
variables of the field in cavity 1 and cavity 2, and, for each $n$ and $m$,
$V_{nm}(t)=V_{nm}(t)^T=V_{mn}^*(t)$ are symmetric and complex four-by-four matrices, $y_{nm}(t)=y_{mn}^*(t)$ are complex four-by-one vectors, and $N_{nm}(t)=N_{mn}^*(t)$ are complex scalars. The problem thus reduces to a finite set of coupled stochastic differential equations for $V_{nm}(t)$, $y_{nm}(t)$, and $N_{nm}(t)$, which may be solved numerically for given realizations of the noise $dW$. In addition, for imaginary $\epsilon$ and $\beta$ and real $e^{i\phi}$ the time-reversal symmetry described above implies that the four-by-four matrices $V_{-m,-n}(t)$ and $V_{nm}(t)$ are related by a change of sign of the elements in the second and third rows followed by a change of sign of the elements in the second and third columns, that $y_{-m,-n}(t)$ is obtained from $y_{nm}(t)$ by changing the sign of the second and third rows, and that $N_{-m,-n}(t)=N_{nm}(t)$ if $N_{-m,-n}(0)=N_{nm}(0)$. As expected, it thus follows that the Wigner function phase space representation of the field in cavity 1 is symmetric under reflection in the $x_1$-axis and that the atomic states $|n\rangle\langle n|$ and $|-n\rangle\langle-n|$ have equal weights at all times if they have equal weights at the initial time.

The differential equations reveal that the evolution of
$V_{nm}(t)$ is deterministic and that $dV_{nm}(t)/dt$ only depends on
$n$, $m$, $V_{nm}(t)$, and the physical parameters of the setup,
except $\beta$. Numerically we observe that $V_{nm}(t)$ quickly
approaches the steady state solution, and it is thus possible to
reduce the number of differential equations significantly if the
transient is negligible. In contrast to the case without squeezing,
$y_{nm}(t)$ depends on the measurement outcome for $\epsilon\neq0$
because the noise of the field leaking out of the cavities is
correlated with the noise of the field inside the cavities, and the
observations lead to a back-action on the mean value of the field
amplitudes for each $n$ as well as an update of the probabilities. In the unobserved case ($\eta=0$), $y_{nn}(t)$ does, however, approach a steady state value, which for $\phi=\pi$, $\beta=i|\beta|$, and $\epsilon=i\textrm{Im}(\epsilon)$ is $[\sqrt{2}\textrm{Re}(\alpha_{\epsilon,n}),
\sqrt{2}\textrm{Im}(\alpha_{\epsilon,n}),0,
-2|\beta|\sqrt{2\kappa_2}/(\kappa_2+\kappa_{\textrm{loss},2}+2\textrm{Im}(\epsilon))]^T$, where $\alpha_{\epsilon,n}$ is given by the right hand side of Eq.~\eqref{alphanss}, except that $\beta$ is replaced by $|\beta|[2\kappa_2/(\kappa_2+\kappa_{\textrm{loss},2}+2\textrm{Im}(\epsilon))-1]$.

\begin{figure}
\includegraphics*[width=\columnwidth]{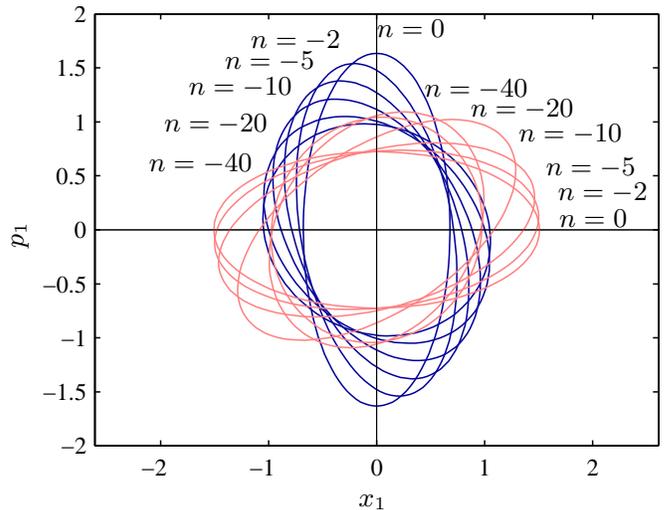}
\caption{(Color online) Steady state uncertainty ellipses of the light field in cavity 1 corresponding to the atomic state $|n\rangle\langle n|$ for
  different values of $n$. The parameters are $g=2\pi\times215$~MHz,
  $\Delta=2\pi\times10$~GHz, $\phi=\pi$, $\epsilon=\pm0.2\textrm{
  }i\kappa_2$, $\kappa_1=\kappa_2=2\pi\times106$~MHz,
  $\kappa_{\textrm{loss},1}=\kappa_{\textrm{loss},2}=0$, and
  $\eta=0.9$. The blue (dark) lines are for $\epsilon=+0.2\textrm{
  }i\kappa_2$ and it is apparent that the state corresponding to $n=0$
  is squeezed in the $x_1$-quadrature and anti-squeezed in the
  $p_1$-quadrature. As $|n|$ increases, the squeezing is rapidly
  reduced and the uncertainty ellipse turns so that the minimal
  (maximal) uncertainty is no longer along $x_1$ ($p_1$). The
  quadratures thus become correlated and back-action due to an $x_1$
  measurement also affects the $p_1$ mean value. The red (bright) lines
  are for $\epsilon=-0.2\textrm{ }i\kappa_2$ and in that case, the
  $n=0$ state is anti-squeezed in the $x_1 $-quadrature and squeezed
  in the $p_1$-quadrature. Note that the error ellipses are
  independent of $\beta$ and that, for the purpose of illustration, we have chosen a rather high value of $|\epsilon|$.}\label{ellipse}
\end{figure}

The stochastic nature of $y_{nn}(t)$ prevents us from drawing a graph
corresponding to Fig.~\ref{alphaplot}, which is valid for all possible
measurement outcomes, and instead we note that $V_{nn}(t)$ is the
four-by-four covariance matrix of the two light field modes
conditioned on the atomic state $|n\rangle\langle n|$ and plot the
corresponding steady state error ellipses of the cavity 1 field in
Fig.~\ref{ellipse}. As in the unsqueezed case, extreme $n$ values,
$|n|\sim J$, leads to cavity 1 being almost in the vacuum state
because the effective resonance frequency of the cavity is shifted by
the atoms and the probe beam is hardly coupled into the cavity at
all. It is thus not surprising that the squeezing is very weak for
these extreme $n$-values, as observed in the figure.

\subsection{Numerical results}
\label{sec:num_results_squeeze}

\begin{figure}
  \includegraphics*[width=0.95\columnwidth]{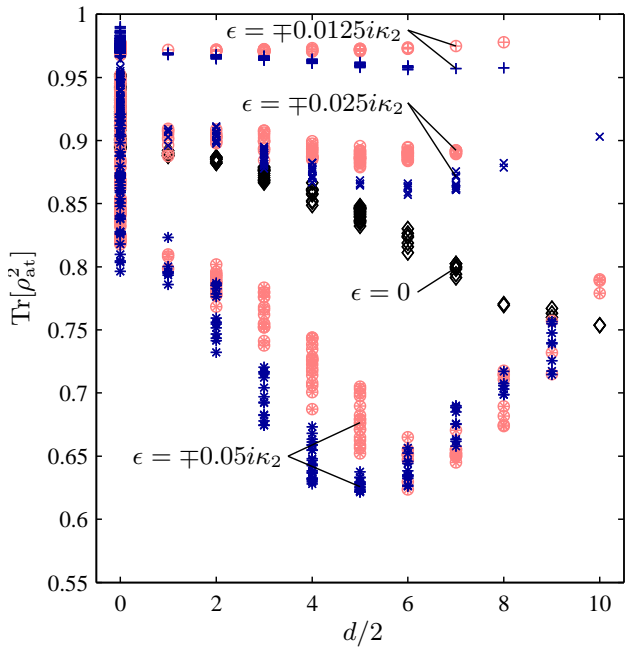}
  \caption{(Color online) Scatter plot of purity vs.\ $d/2$, where $\pm d/2$ are the positions of the peaks of the $n$-distribution of the
    atoms. The larger $d/2$, the more separated are the two
    components of the macroscopic superposition, cf.\
    Fig.~\ref{state}(b). The results are for $J=50$ and coupling and
    loss parameters like in Fig.~\ref{ellipse}. The
    input light and squeezing is on for $t=10^{-6}$~s, but we continue
    observations for an additional time of $15/\kappa_1\sim 2\times
    10^{-8}$~s in order for the cavity light field to decay to
    vacuum. The $\lozenge$ symbols (black) indicate results for a coherent state probe with $4\kappa_1|\beta|^2/\kappa^2=0.01$. The blue (dark) $+$, $\times$, and $*$ symbols indicate results for $\beta=0$ and
    $\epsilon=0.0125,0.025,0.05 \textrm{ } i\kappa_2$, respectively. The red (bright) circled symbols $\oplus$, $\otimes$, and $\circledast$ are
    correspondingly for $\epsilon=-0.0125,-0.025,-0.05 \textrm{ }
    i\kappa_2$. Each series contains 150 points. As can be seen, probing with weakly squeezed vacuum can lead to higher purity than probing with unsqueezed coherent light.}\label{fig:pur_vs_m}
\end{figure}

In Fig.~\ref{fig:pur_vs_m} we plot numerical results for $\beta=0$ and various values of the nonlinear coefficient $\epsilon$ \endnote{We have used a second order derivative free predictor-corrector method, cf.~\cite{kloeden}.}, and, for reference, we also plot results of probing with a coherent state field with $4\kappa_1|\beta|^2/\kappa=0.01$. For nonzero $\epsilon$, the obtained state depends on the whole detection record and every realization of the noise $dW$ leads to a unique state. It is thus more difficult to summarize the results. We have chosen to present a scatter plot of purity versus $d/2$, where
$\pm d/2$ are the positions in the peaks in the $n$ distribution of
the atoms, cf.\ Fig.~\ref{state}(b). Larger values of $d/2$ thus
correspond to a more ``macroscopic'' superposition state.

For the strongest squeezing considered, $\epsilon=\pm0.05\textrm{ }i\kappa_2$, we see that for all but the highest values of $d/2$, the purity is lower than what is obtained with coherent state probing like in Sec.~\ref{III}. We
attribute this to the correlations between the field leaking out of
the cavities and the field inside the cavities: In contrast to a
coherent state, the purity of a squeezed state decreases if it is
subjected to loss, and the stochastic movement of the mean values of
the cavity field for different values of $n$ may increase the distance
in phase space between the cavity fields corresponding to the same
value of $|n|$.

In fact, it is not quite fair to compare the $\epsilon=\pm0.05\textrm{
}i\kappa_2$ results to the coherent state results as for the chosen
observation time, $t=10^{-6}$~s, the former has more peaked $n$
distributions than the latter. This means that one could have reduced
$t$ in the squeezed case. Instead we keep $t$ fixed and look at
results for $\epsilon=\pm 0.025\textrm{ }i\kappa_2$, which for
$t=10^{-6}$~s leads to $n$-distributions very similar to the coherent
state results. As can be seen from Fig.~\ref{fig:pur_vs_m}, we can
then preserve a high purity while still being sensitive enough to make
the $n$ distribution double-peaked. The highest purity is found for
the phase of $\epsilon$ where the $n=0$ light-state is squeezed in the
$p_1$ quadrature. A closer analysis reveals that for higher $|n|$ the
measurement back-action is then less likely to drive the light field
amplitudes for $\pm n$ apart and thus the field leaking to the
environment carries less information on the atomic state.

Finally, we can also consider very weak squeezing, $\epsilon=\pm
0.0125\textrm{ }i\kappa_2$. The purity is then above 95\% for all
$d/2$ for all simulations. For such weak squeezing, however, the
peaks in the $n$-distribution are considerably wider than for the
coherent state scheme. It may be noted that for the very weak
squeezing, the phase of $\epsilon$ apparently has a pronounced effect
on the purity even for the highest values of $d/2$, while for
stronger squeezing, the difference disappears as $d/2$ is increased.

\section{Conclusion}\label{V}

In conclusion, we have analyzed a protocol, which uses optical
measurements to conditionally prepare an ensemble of spin-$1/2$ atoms
in a superposition of a state with most of the atoms in the spin up
state and a state with most of the atoms in the spin down state. The
physics of the protocol may be understood by referring to the plot of
the cavity field in Fig.~\ref{alphaplot}. A large horizontal distance
between two states means that the states are easy to distinguish in an
$x$-quadrature measurement, and the vertical distance between the
states $n$ and $-n$ determines the rate of decrease of purity due to
light field losses. Increasing $2\tilde{g}/\kappa$ moves the coherent
states around the edge of a circle towards the origin, and for an
initially coherent spin state, this increases the purity obtained
after a given probing time and a favorable measurement outcome, but it
also decreases the distance $d$ between the two peaks obtained in the
distribution over the different $\hat{J}_z$ eigenstates, because for
large $2\tilde{g}/\kappa$ only states with $n$ very close to zero are
easily distinguished from states with $n$ close to $\pm J$. This is,
however, of no concern for $J=1$, and in this case perfect pure
quantum superposition states are obtained in the limit
$2\tilde{g}/\kappa\rightarrow\infty$. An increased number of atoms
increases $d$ and the purity of the final states, because more states
are added to the left in Fig.~\ref{alphaplot}, but the increase in $d$
is small, and $d/N$ decreases with $N$. Both the rate of gain of
information, the rate of loss of photons, and the rate of spontaneous
emission scale linearly with the probe beam intensity.

We have also investigated probing with a continuous beam of squeezed vacuum. This system is significantly harder to analyze because the back-action effect of the measurements is more complicated, but we have found numerically that it is possible to achieve a higher purity of the generated states if the probe beam is slightly squeezed. The stochastic master equation \eqref{sqrho} is quite general and may be used to analyze probing with squeezed light in other settings as well.

The phase shift of the light field per round trip in the cavity has been assumed to be infinitesimally small, but we note that
additional possibilities arise if the phase shift per round trip is
comparable to $2\pi$. A pure superposition of $|J\rangle$ and
$|-J\rangle$ can, for instance, be obtained if the phase shift per
round trip is $\pm\pi$ for $n=\mp J$ and light field losses within the
atomic ensemble are negligible, since, in that case, there is no
difference between the cavity field strength for the $n=\pm J$ atomic states, but only a difference between the local phase variation of the fields inside the atomic medium. Similarly, if the phase shift is $\pi$ per round
trip per atom and $N$ is even, it is possible to generate pure
superpositions of states with even values of $n$ and pure
superpositions of states with odd values of $n$.

A coherent spin state is a convenient starting point, because it has the desired initial symmetry and is relatively easy to prepare experimentally, but it could also be interesting to consider other initial states. In \cite{polzik}, it has, for instance, been suggested to spin squeeze the coherent spin state in order to increase the weight of the larger $|n|$ components. Another interesting possibility is to start from a time-reversal eigenstate. For such a state the additional spin symmetries in $H$ are not required, cf.\ Sec.~\ref{sec:time-reversal}, and it is possible to perform arbitrary spin rotations during the probing without breaking the $\pm n$ symmetry. This opens the way to use a feedback scheme similar to the scheme analyzed in \cite{stockton} to force the atoms into an equal superposition of $J$ and $-J$ (for the case where the initial state is an eigenstate of ${\bm\hat{J}}^2$ with eigenvalue $\hbar^2J(J+1)$). Note that $|0\rangle\langle0|$ (and any rotated version of this state) fulfills these requirements.

\begin{acknowledgments}
This work was supported by the European Commission through the Integrated Project FET/QIPC SCALA.
\end{acknowledgments}

\end{document}